\documentclass[conference]{IEEEtran}
\IEEEoverridecommandlockouts
\usepackage{flushend}
\usepackage{caption}
\usepackage{subcaption}
\usepackage{graphicx}
\usepackage[table, svgnames, dvipsnames]{xcolor}
\usepackage{booktabs}
\usepackage{cite}
\usepackage{amsmath,amssymb,amsfonts}
\definecolor{green}{rgb}{0.16, 0.67, 0.53}
\usepackage{subcaption}
\usepackage{tabularx}
\usepackage{mathtools}
\usepackage{amsmath}
\usepackage{blkarray}
\usepackage{rotating}
\usepackage{svg}
\usepackage{enumerate}
\usepackage{enumitem}
\usepackage{algorithmic}
\usepackage{algorithm}
\usepackage{color}
\usepackage{xcolor}
\usepackage{multirow}
\usepackage{lineno}
\usepackage[export]{adjustbox}
\usepackage[utf8]{inputenc}
\usepackage{makecell}
\usepackage{multicol}
\usepackage{pifont}
\usepackage[hidelinks]{hyperref}

\definecolor{mypink3}{cmyk}{0, 0.47808, 0.10429, 0.012}

\usepackage{textcomp,balance,url}
\usepackage[left=1.62cm,right=1.62cm,top=1.9cm]{geometry}
\definecolor{chestnut}{rgb}{0.97, 0.51, 0.47}
\def\BibTeX{{\rm B\kern-.05em{\sc i\kern-.025em b}\kern-.08em
    T\kern-.1667em\lower.7ex\hbox{E}\kern-.125emX}}

\definecolor{blue}{rgb}{0.0, 0.47, 0.75}

\usepackage{soul}                                             \sethlcolor{yellow}

\begin{document}
\title{An Uncertainty-Aware Resilience Micro-Agent for Causal Observability in the Computing Continuum}

\author{
\IEEEauthorblockN{Suvi De Silva\IEEEauthorrefmark{3}, Alfreds Lapkovskis\IEEEauthorrefmark{3}, Alaa Saleh\IEEEauthorrefmark{2}, Sasu Tarkoma\IEEEauthorrefmark{2}, and Praveen Kumar Donta\IEEEauthorrefmark{3}}

\IEEEauthorblockA{\textit{\IEEEauthorrefmark{3}Department of Computer Systems and Sciences}, \textit{Stockholm University,} Stockholm 164 25, Sweden\\  \texttt{suvi.desilva@outlook.com, \{alfreds.lapkovskis, praveen\}@dsv.su.se}} 
\IEEEauthorblockA{\IEEEauthorrefmark{2}\textit{Department of Computer Science}, \textit{University of Helsinki,} Helsinki 00014, Finland}
 \texttt{\{alaa.saleh, sasu.tarkoma\}@helsinki.fi}
 }
 \maketitle

\begin{abstract}
Grey failures in the computing continuum produce ambiguous overlapping symptoms that existing approaches fail to diagnose reliably, either due to a lack of causal awareness or acting under high epistemic uncertainty, risking destructive interventions. This paper presents an uncertainty-aware resilience micro-agent for causal observability (AURORA), a lightweight framework for diagnosing and mitigating grey failures in edge-tier environments. The framework employs parallel micro-agents that integrate the free-energy principle, causal do-calculus, and localized causal state-graphs to support counterfactual root-cause analysis within each fault’s Markov blanket. Restricting inference to causally relevant variables reduces computational overhead while preserving diagnostic fidelity. AURORA further introduces a dual-gated execution mechanism that authorizes remediation only when causal confidence is high and predicted epistemic uncertainty is bounded; otherwise, it abstains from local intervention and escalates the diagnostic payload to the fog tier. Our experiments demonstrate that AURORA outperforms baselines, achieving a 0\% destructive action rate, while maintaining 62.0\% repair accuracy and a 3ms mean time to repair.
\end{abstract}

\begin{IEEEkeywords}
Active Inference, Bayesian Networks, Computing Continuum, Micro-Agent, Resilience
\end{IEEEkeywords}

\section{Introduction}
\IEEEPARstart{C}{omputing} continuum enables services to place computation closer to data sources while retaining access to higher-capacity processing resources when needed.
However, heterogeneity also complicates runtime resilience, as faults may arise from interactions among constrained compute resources, variable network conditions, and tier-dependent orchestration policies \cite{khan2026governance}. A major challenge in such environments is the occurrence of gray failures \cite{donta2025resilient}, including network congestion, CPU contention, and memory leakage, which can produce overlapping symptoms such as throughput degradation, increased latency, and service-level objective (SLO) violations. Because these symptoms can also arise from transient workload fluctuations, reliable diagnosis requires causal reasoning rather than relying solely on threshold-based or correlation-driven monitoring.

Research on resilience in the computing continuum is still emerging; accordingly, this section reviews the most recent and closely related works. Donta et al. \cite{donta2023governance} introduced a big data approach to the governance and sustainability of distributed continuum systems. While providing a robust, high-level control plane for data lifecycle management, this macro-level governance approach faces challenges in executing real-time, autonomous triage natively on strictly resource-constrained edge devices. Further, \cite{rouska2026equilibrium,donta2025resilient} construct causal fault graphs from operational logs and manage epistemic uncertainty through Markov blankets, demonstrating the feasibility of active inference (AIF) on resource-constrained edge hardware. Decentralized multi-agent frameworks such as Symphony \cite{wang2025symphony} further show that lightweight coordination can be distributed across edge nodes to reduce single-node processing bottlenecks. However, these approaches do not explicitly regulate remediation under diagnostic uncertainty. To address scalability, Saleh et al. \cite{saleh2026bio} proposed a bio-inspired agentic self-healing framework that provides an organically adaptable topology. Similarly, Ye et al. \cite{ye2026nesy} advanced the domain by proposing a neuro-symbolic approach to trustworthy self-healing, enhancing logical rule mapping to neural observations. While highly innovative, these architectures are computationally heavy and lack explicit mathematical safety bounds that prevent an agent from acting when epistemic uncertainty is dangerously high.

\begin{table}[htbp]
\centering
\scriptsize
\caption{Related works.}
\label{tab:related_works}
\renewcommand{\arraystretch}{1.3}
\begin{tabularx}{\linewidth}{@{} l >{\raggedright\arraybackslash}X >{\raggedright\arraybackslash}X >{\raggedright\arraybackslash}X @{}}
\toprule
\textbf{Ref.} & \textbf{Method} & \textbf{Benefit} & \textbf{Gaps} \\
\midrule
\cite{donta2023governance} & Big data governance & Robust data lifecycle control & Macro-level; lacks edge-native triage. \\
\cite{sedlak2024equilibrium} & BNSL + Markov blankets & Reduces edge compute overhead & No active execution or safety gates. \\
\cite{sedlak2025adaptive} & AIF stream processing & Localized bottleneck detection & Missing parallel causal diagnostics. \\
\cite{donta2025resilient} & PAIR-agent (AIF) & Maps causal fault graphs on edge & Operates without VFE safety limits. \\
\cite{wang2025symphony} & Multi-agent scheduling & Bypasses edge resource limits & Lacks causal diagnostics / VFE bounds. \\
\cite{lapkovskis2025benchmarking} & Benchmarking AIF vs. reinforcement learning & Evaluates compliance metrics & Shows need for safe action baselines. \\
\cite{ye2026nesy} & NeSy self-healing & Logical rule mapping & Computationally heavy for edge nodes. \\
\cite{saleh2026bio} & Bio-inspired continuum & Scalable organic topology & Lacks formal bounds on execution. \\
\bottomrule
\end{tabularx}
\par\raggedright
\end{table}

Further, rule-based policies and static thresholds map symptoms to predefined remediation actions \cite{abreu2025resilience,parashar2025autonomic}. However, without causal representation, they cannot reliably distinguish among competing fault hypotheses, leading to remediation being triggered under uncertainty and potentially missing the true root cause.  In grey failure scenarios, incorrect interventions may restart healthy services, obscure faults through threshold changes, or propagate contention through workload redistribution. This limitation is especially critical at the edge, where constrained CPU, memory, and energy limit the use of heavyweight diagnostic models, motivating lightweight, uncertainty-aware resilience mechanisms \cite{sen2024fault}. 
Table~\ref{tab:related_works} summarizes the key gaps identified across related works.

In this context, we present \textit{an uncertainty-aware resilience micro-agent for causal observability} (AURORA), a lightweight framework for failure diagnosis and mitigation at the edge.
The main technical contributions are summarized as follows:
\begin{itemize}
    \item AURORA defines a lightweight parallel micro-agent architecture for grey failure management, in which agents perform localized causal observability and AIF-based recovery reasoning under posterior fault belief and epistemic uncertainty.
    \item It constructs causal models using Bayesian networks (BNs) with Markov blanket-constrained inference to enable bounded-complexity, real-time counterfactual root-cause diagnosis via do-calculus.
    \item AURORA introduces a dual-gated execution mechanism that authorizes local recovery only when the inferred causal hypothesis satisfies a posterior confidence constraint, and the candidate intervention satisfies a bounded variational-free-energy (VFE) criterion; otherwise, unresolved cases are escalated to the fog tier.
\end{itemize}

The remainder of this paper is organized as follows. Section~\ref{sec:Background} presents the preliminary technical foundations of AURORA. Section~\ref{sec:architecture} describes the proposed AURORA framework. Section~\ref{sec:Results} presents the results and discussion. Finally, Section~\ref{sec:Conclusion} concludes the paper and outlines future directions.

\section{Preliminaries}\label{sec:Background}
This section summarizes AURORA's technical foundations. Some design choices follow Sedlak et al.~\cite{sedlak2024equilibrium}.
\subsection{Bayesian Networks and Posterior Queries}
A BN~\cite{darwiche2008bayesian} is a directed acyclic graph (DAG) in which nodes represent random variables, and edges encode conditional dependencies. Given observed evidence $X$, posterior inference over a target variable $Y$ follows 
\begin{equation}\label{eq:posterior_pre}
  P(Y \mid X) \propto P(X \mid Y)\,P(Y).
\end{equation}
In practice, posterior queries are computed via variable elimination, which marginalizes out non-target variables in a fixed elimination order~\cite{darwiche2008bayesian}.

\subsection{Structure Learning and BIC}
BN structure can be learned from data using Bayesian network structure learning (BNSL)~\cite{scutari2019learning}. AURORA uses hill-climbing search scored by the Bayesian information criterion (BIC), which balances goodness-of-fit against model complexity to prevent overfitting on data-sparse edge devices~\cite{sedlak2024equilibrium}.

\subsection{Markov Blankets}
The Markov blanket $\mathcal{MB}(v)$ of a node $v$ consists of its parents, children, and co-parents in the DAG, and satisfies:
\begin{equation}
  P\!\left(v \mid \mathcal{V} \setminus v\right) = P\!\left(v \mid \mathcal{MB}(v)\right).
  \label{eq:mb_pre}
\end{equation}
This conditional independence allows posterior inference over $v$ to be restricted to $\mathcal{MB}(v)$ without loss of accuracy~\cite{aliferis2010local,gao2016efficient}, making real-time causal inference tractable on resource-constrained devices.

\subsection{Pearl's do-Calculus}
Pearl's do-calculus~\cite{pearl2009causality} formalizes intervention reasoning. The operator $\mathrm{do}(X=x)$ sets $X$ to $x$ by severing its incoming DAG edges, thereby blocking non-causal paths:
\begin{equation}
  P\!\left(Y=y \mid \mathrm{do}(x)\right)
  \label{eq:do_pre}
\end{equation}
estimates the probability of $Y=y$ \emph{if} action $x$ is applied, without executing it on the live system. This is the quantity that AURORA's safety gate evaluates before dispatching any remediation.

\subsection{Active Inference and Variational Free Energy}
AIF~\cite{friston2010free} is a neuroscience-grounded framework in which an agent minimizes its VFE, a measure of surprise between its generative model and observations. Given approximate posterior $Q(s)$ over latent states $s$ and generative model $P(o,s)$ with observations $o$, VFE is defined as
\begin{equation}
  \mathcal{F} \triangleq \mathbb{E}_{Q(s)}\!\left[\ln Q(s) - \ln P(o, s)\right].
  \label{eq:vfe_pre}
\end{equation}
Low VFE indicates that the agent's model is consistent with current observations; high VFE signals epistemic surprise. AURORA uses $\mathcal{F}$ as a second safety gate: an action is authorized only when its predicted post-intervention VFE remains below a threshold, ensuring the model is calibrated before any physical remediation is executed. For the full AIF cycle and its application to edge SLO enforcement, see~\cite{sedlak2024equilibrium}.

\section{Proposed AURORA Framework}
\label{sec:architecture}
\begin{figure}[!t]
\centerline{\includegraphics[width=\columnwidth]{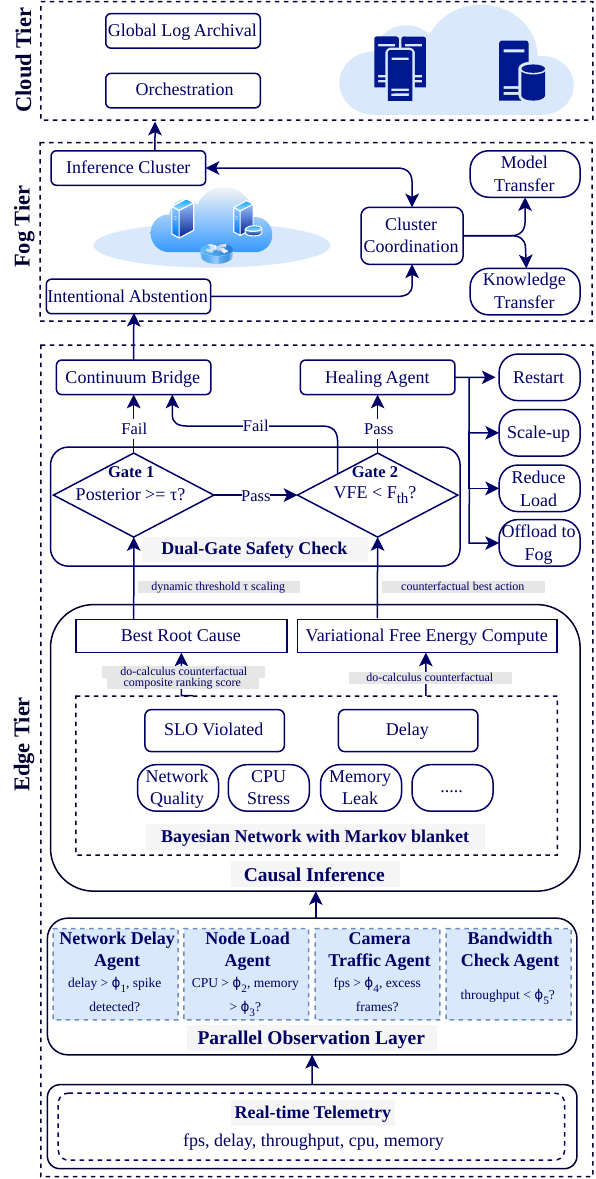}}
\caption{AURORA micro-agent pipeline architecture for the computing continuum.}
\label{fig:cc}
\end{figure}

The design of AURORA is motivated by a fundamental tension in edge computing. Edge devices must diagnose and resolve faults autonomously, without incurring the round-trip latency of a cloud consultation, yet their constrained CPU and memory budgets preclude the exhaustive inference models that reliable autonomous action requires. 
AURORA addresses this gap through three coordinated design principles: (i) compact causal models that fit the resource-constrained devices while supporting counterfactual intervention reasoning; (ii) a parallelized, single-responsibility micro-agent pipeline that maximizes observational coverage without creating processing bottlenecks on edge threads; (iii) and a dual-gated execution mechanism that checks the confidence of the inferred root cause and the uncertainty of the proposed remediation before taking action. If either condition is not satisfied, the micro-agent abstains from local remediation and forwards the case for deeper analysis. Fig.~\ref{fig:cc} illustrates how these principles are distributed across the three tiers of the computing continuum. Diagnostic autonomy is decentralized to the edge tier, where micro-agents perform causal inference directly on live telemetry. When local recovery is probabilistically unviable, as determined by the safety gates, the continuum bridge serializes the diagnostic payload and forwards it to a fog-tier node for deeper analysis. The following subsections describe each component in detail.

\subsection{Causal Models}
\label{subsec:causal_models}
AURORA requires a diagnostic model that is lightweight, probabilistic, and intervention-aware. The model must run on resource-constrained edge devices, estimate posterior fault probabilities for the safety gates, and support queries about the effect of candidate remediation actions. BNs~\cite{darwiche2008bayesian} satisfy these requirements and are therefore used as the causal backbone of AURORA.

Concretely, AURORA represents the runtime system state as a BN over telemetry variables, including CPU utilization, memory usage, network throughput, delay, and a binary SLO outcome node $S$. Posterior root-cause queries $P(C \mid O_t)$ are evaluated over candidate causes $C$ using variable elimination restricted to the Markov blanket $\mathcal{MB}(S)$ in Eq.~\eqref{eq:mb_pre}, which keeps inference tractable on Raspberry Pi-class edge hardware. Interventional queries $P(S=1 \mid \mathrm{do}(a))$ in Eq.~\eqref{eq:do_pre} are evaluated by graph mutilation, where incoming edges to the intervened variable are removed, and the required adjustment is applied internally.

AURORA's causal graph is pre-specified from domain knowledge and held fixed across trials; only the conditional probability tables are fitted from a held-out training sample. While dynamic structure and parameter learning were investigated as candidate extensions, they were disabled for the reported experiments, as greedy search on small synthetic samples can silently introduce structural errors that can mislead root-cause analysis.

Each recovery action is represented as a structural patch, i.e., the set of causal state variables whose values are fixed by the intervention and passed to $\mathrm{do}(\cdot)$. Table~\ref{tab:action_patches} summarizes these patches; for example, $\mathrm{do}(\texttt{restart})$ fixes $\{\texttt{cpu}=\text{normal},\,\texttt{delay}=\text{normal}\}$ in the BN. This mapping allows Gate~2 in Section~\ref{subsec:dual_gated} to compute the expected SLO state under each candidate recovery action before applying the action in the real-time system.
\begin{table}[t] 
  \caption{Action$\to$Patch Map: structural intervention applied by $\mathrm{do}(a)$.}                            
  \label{tab:action_patches}                
  \centering                 
  \renewcommand{\arraystretch}{1.15}
  \begin{tabular}{@{}ll@{}}          
  \toprule         
  \textbf{Action $a$} & \textbf{Patch applied by $\mathrm{do}(a)$} \\       
  \midrule         
  \texttt{restart}          & $\{\texttt{cpu}{=}\text{normal},\, \texttt{delay}{=}\text{normal}\}$ \\ 
  \texttt{scale\_up}        & $\{\texttt{memory}{=}\text{normal},\, \texttt{cpu}{=}\text{normal}\}$ \\
  \texttt{offload\_to\_fog} & $\{\texttt{network\_quality}{=}\text{high},\, \texttt{delay}{=}\text{normal}\}$ \\  
  \texttt{reduce\_load}     & $\{\texttt{cpu}{=}\text{normal},\, \texttt{delay}{=}\text{normal}\}$ \\ 
  \bottomrule                 
  \end{tabular}                                           
\end{table}

\subsection{Parallel Diagnostic Pipeline}
\label{subsec:diagnostic_pipeline}

Given the causal backbone described above, AURORA executes diagnosis and recovery through an ordered micro-agent pipeline. Observation agents operate in parallel over disjoint system facets, while downstream inference and safety checks proceed sequentially before any recovery action is executed. Fig.~\ref{fig:cc} illustrates the pipeline transitions, which consist of the following phases.
\subsubsection{Parallel Diagnostic Monitoring}
AURORA begins with telemetry ingestion. At each observation step, the telemetry vector $O_t$ is passed to a coordination agent, which dispatches monitoring tasks to four concurrent single-responsibility micro-agents. Each micro-agent evaluates a distinct SLO predicate $\phi_i$ over a specific system facet, as shown in Fig.~\ref{fig:cc}:
  \begin{itemize}
    \item \emph{Network Delay Agent:} flags when end-to-end delay exceeds $\phi_1$~ms.
    \item \emph{Node Load Agent:} flags when CPU utilization exceeds $\phi_2$ or memory utilization exceeds $\phi_3$.
    \item \emph{Camera Traffic Agent:} flags when per-camera frame rate exceeds $\phi_4$~fps, (from our use case in the experiments).
    \item \emph{Bandwidth Check Agent:} flags when throughput drops below $\phi_5$~MB/s.
  \end{itemize}
Partitioning telemetry into four facets limits each micro-agent to the metrics required for its SLO predicate, reducing per-agent state and avoiding a serial monitoring bottleneck on the edge device. It also decouples the thresholds $\phi_i$, allowing each predicate to be tuned independently without affecting the remaining detection logic. When a predicate is violated, the corresponding micro-agent emits an anomaly flag. The coordination agent aggregates these flags into an anomaly vector $A_t \subseteq O_t$ and forwards it to the next stage only if $A_t \neq \emptyset$, thereby bypassing causal inference during nominal operation.

\subsubsection{Causal Inference and Root Cause Extraction}
Let $S\in\{0,1\}$ denote a binary SLO random variable, where $1$ indicates fulfillment and $0$ violation. Upon receiving anomaly vector $A_t \neq \emptyset$, a dedicated \emph{inference agent} first queries the Markov-blanket-bounded BN $\mathcal{B}$ for the posterior distribution over candidate root causes $C \in \mathcal{MB}(S)$, as shown in Eq.~(\ref{eq:posterior}),
\begin{equation}
    P(C \mid O_{t}) \propto P(O_{t} \mid C)\, P(C).
    \label{eq:posterior}
\end{equation}
Markov blanket restriction makes variable elimination feasible on resource-constrained devices, since nodes outside the blanket are conditionally independent of the symptom, as defined in Eq.~\eqref{eq:mb_pre}, and can be excluded without loss of diagnostic accuracy. 

Although the Markov blanket posterior provides a tractable estimate of candidate root causes, posterior probability alone can be insufficient for ranking faults under grey-failure conditions. Multiple SLO predicates may degrade concurrently, and telemetry variables may remain correlated across fault classes, causing the posterior to favor a plausible but incorrect hypothesis, such as attributing a delay spike to CPU contention when the actual cause is bandwidth degradation. To reduce this correlation-induced ambiguity, AURORA ranks each candidate cause using an \emph{interventional contrast}. This score measures the change in symptom probability between the observed state and a benign counterfactual assignment of $C$ under the do-operator. Since the intervention severs incoming causal links to $C$, the score suppresses associations induced by common ancestors and emphasizes the causal effect of $C$ on the observed symptom. The interventional contrast is defined as     
\begin{equation}
  \begin{aligned}                    
    \Delta(C) \triangleq \big|&P(S=0 \mid \mathrm{do}(C=v_{\text{obs}}),Z) \\
    -&P(S=0 \mid \mathrm{do}(C=v_{\text{benign}}),Z) \big|,
  \end{aligned}  
  \label{eq:impact}
  \end{equation}          
where $Z$ is the conditioning set obtained by removing $C$, its descendants, and $S$ from $O_t$. Conditioning on a descendant of the intervened variable would open a non-causal path and bias the contrast, so descendant pruning of $Z$ is required for the contrast to be a valid causal-effect estimate. Each candidate cause is then assigned a composite score $\rho(C)$ that combines interventional contrast with supporting diagnostic evidence. This score integrates the estimated causal effect of $C$ on the symptom with corroborating posterior and telemetry-based signals, providing a more stable ranking under correlated grey-failure observations, as shown in Eq.~\eqref{eq:ranking}         
\begin{equation}
  \rho(C) \triangleq w_1\,\Delta(C) + w_2\,\sigma(C)
           + w_3\,\mathbf{1}\left[C \in \mathrm{Pa}(S)\right]+r(C)
    \label{eq:ranking}
  \end{equation}
where $\Delta(C)$ denotes the interventional contrast, $\sigma(C)\in[0,1]$ is the normalized observed severity of candidate cause $C$, $\mathbf{1}\left[C \in \mathrm{Pa}(S)\right]$ assigns an additional weight to direct parents of the symptom node $S$, and $r(C)$ is structural prior $r(C)$ that assigns bonus for root nodes in $\mathcal{B}$ and penalty for mediator nodes, reflecting the assumption that true fault causes are more likely to originate at source nodes than along intermediate causal paths. The weights $w_1+w_2+w_3=1.0$ control the relative contributions of causal contrast, observed severity, and direct-parent evidence, respectively. The top-ranked cause is selected as
$C_{\text{best}} \triangleq \arg\max_C \rho(C)$, and its posterior probability $P_{\max}$ is forwarded to the dual-gated safety evaluation.

\subsection{Dual-Gated AIF and Dynamic Scaling}
\label{subsec:dual_gated}
The dual-gated design is motivated by the distinction between diagnostic confidence and intervention reliability. Existing AIF-based frameworks~\cite{sedlak2024equilibrium,donta2025resilient} typically authorize execution using the maximum posterior probability $P_{\text{max}}$, allowing the agent to act when the inferred root cause is sufficiently probable. Although this tests diagnostic certainty, it does not verify whether the generative model can reliably predict the effect of the selected action under the current system state. In grey-failure conditions, posterior belief may concentrate on a plausible cause even when the model remains poorly calibrated to the observed dynamics, increasing the risk of unsafe recovery actions. AURORA addresses this limitation through a second gate based on VFE. Gate~1 verifies that the diagnosis is sufficiently confident, while Gate~2 verifies that the predicted intervention remains consistent with the current observations. The complete procedure is formalized in Algorithm~\ref{alg:aurora}.
\begin{algorithm}[t]
    \caption{AURORA Dual-Gated Active Inference Loop} 
    \label{alg:aurora}
    \begin{algorithmic}[1]
    \REQUIRE Observation $O_t$; Bayesian Network $\mathcal{B}$;
             \\gate parameters $\tau_{\text{base}}$, $\tau_{\min}$,  
             $\lambda$, $\mathcal{F}_{\text{th}}$
    \ENSURE Mitigation action $a^{*}$ or abstention signal $\bot$                          
    \IF{$O_t$ shows no SLO violation}  
        \STATE \textbf{return} \textsc{healthy} 
    \ENDIF                      
    \STATE \textit{// Stage 1: Parallel Monitoring \& Posterior Ranking}
    \STATE $A_t \leftarrow \texttt{ParallelMonitor}(O_t,\, \{\phi_i\}_{i=1}^{5})$
           \COMMENT{anomaly vector}
    \STATE $P(C_i \mid O_t) \propto P(O_t \mid C_i)\,P(C_i),\;
            \forall C_i \in \mathcal{MB}(S)$ 
           \COMMENT{Eq.~\eqref{eq:posterior}, exact via Variable Elimination}            
    \STATE $C_{(1)} \leftarrow \arg\max_{i}\rho(C_i)$  
           \COMMENT{top-ranked cause, Eq.~\eqref{eq:ranking}}  
    \STATE $P_{\max} \leftarrow P(C_{(1)} \mid O_t)$ 
    \STATE \textit{// Stage 2: Action-Conditional Free-Energy Decomposition} 
    \STATE $a \leftarrow \pi(C_{(1)})$ 
           \COMMENT{policy lookup; patch from Table~\ref{tab:action_patches}} 
    \STATE $(\mathcal{F}_{\text{pragmatic}},\, \mathcal{F}_{\text{epistemic}})           
            \leftarrow \texttt{VFEDecompose}(\mathcal{B},\, O_t,\, a)$   \COMMENT{Eqs.~\eqref{eq:pragmatic},~\eqref{eq:epistemic}}                                 
    \STATE $\mathcal{F} \leftarrow \mathcal{F}_{\text{pragmatic}} + \mathcal{F}_{\text{epistemic}}$                      
    \STATE \textit{// Stage 3: Dynamic Threshold Scaling} 
    \STATE $\tau \leftarrow \max\!\left(\tau_{\min},\;  
            \tau_{\text{base}} - \lambda \cdot \min(|A_t|-1,\,2)\right)$ 
           \COMMENT{Eq.~\eqref{eq:scaling}} 
    \STATE \textit{// Stage 4: Dual-Gated Safety Check} 
    \IF{$P_{\max} < \tau$}  
        \STATE \textbf{return} $\bot_{\text{low-certainty}}$ 
               \COMMENT{Posterior Certainty Gate $\rightarrow$ fog tier} 
    \ELSIF{$\mathcal{F} \geq \mathcal{F}_{\text{th}}$}
        \STATE \textbf{return} $\bot_{\text{high-VFE}}$ 
            \COMMENT{VFE Safety Gate $\rightarrow$ fog tier}
    \ELSE 
        \STATE $a^{*} \leftarrow a$  
               \COMMENT{both gates passed}
        \STATE \texttt{Execute}($a^{*}$ on affected node)
        \STATE \textbf{return} $a^{*}$ 
    \ENDIF 
    \end{algorithmic} 
  \end{algorithm}   

\subsubsection{Dual-Gated Safety Mechanism}
The two gates are deliberately chosen to enforce orthogonal safety properties. Gate~1 is a \emph{diagnostic-ambiguity} test that operates on the posterior over causes, $P(C \mid O_t)$; it can be satisfied by any sharply concentrated posterior. Gate~2 is a \emph{model-consistency} test that operates on the post-intervention generative model; it can be satisfied only when the proposed action's predicted effects are consistent with the current observations. Either failure mode alone is sufficient to trigger a destructive intervention: a concentrated but incorrect posterior may bypass Gate~1, while an inconsistent generative model may bypass Gate~2. Therefore, both gates must pass for execution to proceed.
  
  The \textbf{Posterior Certainty Gate} (Gate~1) enforces $P_{\max} \geq \tau$, where $\tau$ is the certainty threshold with default value $\tau_{\text{base}}$. When the maximum posterior falls below this threshold, the posterior mass is spread across multiple root-cause hypotheses, indicating that the agent cannot reliably distinguish which fault class it is observing. Because incorrect mitigations in a heterogeneous multi-tier environment frequently amplify the original fault rather than resolving it, Gate~1 treats diagnostic ambiguity as a sufficient reason for abstention. 
  
  The \textbf{VFE Safety Gate} (Gate~2) addresses a failure mode that Gate~1 cannot catch: a confident diagnosis whose predicted effects under action are inconsistent with the observed telemetry. This gate computes VFE $\mathcal{F}$.
  For the gate to be operationally useful on edge hardware, AURORA decomposes $\mathcal{F}$ into two additive terms $\mathcal{F}_{\text{pragmatic}}$ and $\mathcal{F}_{\text{epistemic}}$, defined as
\begin{align}                                               \mathcal{F}_{\text{pragmatic}} &\triangleq
  -\log P\!\left(S=1\mid \mathrm{do}(a),\, \mathcal{E}\right),
    \label{eq:pragmatic} \\
    \mathcal{F}_{\text{epistemic}} &\triangleq    
      \frac{1}{|\mathcal{V}'|}      
      \sum_{v \in \mathcal{V}'}
      D_{\text{KL}}\!\left[P_{\text{post}}(v) \,\|\, P_{\text{pre}}(v)\right],
    \label{eq:epistemic}
  \end{align}           
where $a$ is a mitigation action, $\mathcal{E}$ is the current evidence with $S=0$ and the intervened variable removed, $\mathcal{V}'$ is the set of non-target nodes in $\mathcal{B}$, and $P_{\text{post}}, P_{\text{pre}}$ are the predictive marginals after and before applying the action's patch from Table~\ref{tab:action_patches}. $\mathcal{F}_{\text{pragmatic}}$ penalizes actions whose predicted post-intervention SLO recovery probability is low, while $\mathcal{F}_{\text{epistemic}}$ penalizes actions that destabilize the model's posterior across the rest of the network, regardless of whether they are individually beneficial. The total VFE score is defined as 
  \begin{equation}
      \mathcal{F} \triangleq \mathcal{F}_{\text{pragmatic}} + \mathcal{F}_{\text{epistemic}}.
  \end{equation}

If $\mathcal{F} < \mathcal{F}_{\text{th}}$, the candidate intervention is treated as epistemically admissible, and execution proceeds. The threshold is selected based on the separation observed between the VFE distributions of safe and unsafe recovery outcomes during simulation, so that actions with high predictive inconsistency are rejected. Otherwise, AURORA abstains from local execution. This VFE-based gate is the primary mechanism that prevents destructive interventions.

\subsubsection{Dynamic Threshold Scaling}
A fixed certainty threshold introduces a practical problem in severe multi-fault scenarios. When several SLOs are simultaneously violated, as frequently occurs during cascading grey failures, the posterior distribution over root causes is naturally flatter: multiple concurrent anomaly signals are consistent with several overlapping fault hypotheses, so no single hypothesis accumulates the posterior mass required by $\tau_{\text{base}}$. Applying the base threshold rigidly in this regime would cause AURORA to abstain from nearly every multi-fault scenario, including cases where one root cause is clearly dominant and partial intervention is operationally necessary to prevent total service collapse.
To address this, AURORA relaxes the Gate~1 certainty threshold as a bounded linear function of the anomaly count $n$:
\begin{equation}
    \tau = \max\!\left(\tau_{\min},\; \tau_{\text{base}} - \lambda \cdot \min(n - 1,\, 2)\right),
    \label{eq:scaling}
\end{equation}
where $\tau_{\text{base}}$ is the base threshold value, $\tau_{\min}$ is the minimal threshold, and $\lambda$ is a constant that weights the impact of anomaly count. 
The linear schedule is used because posterior concentration decreases approximately linearly as concurrent anomalies increase in our simulations, making Eq.~\eqref{eq:scaling} consistent with the observed uncertainty trend. The function is also monotone, bounded, and simple to audit on edge hardware, which is important when the threshold controls recovery execution. The clipping term $\min(n-1,2)$ limits relaxation to two additional anomalies, so $\tau$ reaches its lower bound for $n \geq 3$. The lower bound $\tau_{\min}$ preserves a minimum confidence requirement, ensuring that recovery is not authorized when the posterior remains broadly ambiguous. Importantly, Gate~2 is not relaxed with $\tau$. Even under multiple concurrent anomalies, a candidate recovery action is executed only if it satisfies $\mathcal{F} < \mathcal{F}_{\text{th}}$, ensuring that increased responsiveness does not bypass the model-consistency check.

\subsubsection{Mitigation or Intentional Abstention}
If both safety gates are satisfied (Algorithm~\ref{alg:aurora}, line~21), the mitigation agent selects the action that maximizes the do-calculus posterior of SLO recovery~\cite{pearl2009causality},
  \begin{equation*}
    a^{*} = \arg\max_{a}\, P\!\left(S=1 \mid \mathrm{do}(a)\right),     
  \end{equation*}
  where $\mathrm{do}(a)$ applies the structural patch listed in Table~\ref{tab:action_patches}, and dispatches the corresponding remediation script. Conversely, if either safety constraint is violated, the framework formally engages the \emph{abstention paradigm}: the edge node halts local execution, serializes the diagnostic payload $\langle O_t,\, P(C \mid O_t),\, C_{\text{best}},\, \mathcal{F} \rangle$, and offloads it via the continuum bridge to a fog-tier orchestration node~\cite{saleh2026bio,donta2025resilient}. This structured handoff reflects a deliberate design choice: rather than treating uncertainty as an error state to be suppressed, AURORA treats it as a signal that the diagnostic problem exceeds the local capacity of the edge tier, and routes it to the tier best equipped to resolve it. The result is a graceful degradation of autonomy across the continuum, rather than a binary choice between blind local action and high-latency cloud escalation. The design and evaluation of fog-tier re-inference from the escalated payload, as well as the cloud-tier orchestration and global-log-archival layers depicted in Fig.~\ref{fig:cc}, are out of scope for this paper and constitute a natural direction for future work. The present contribution evaluates AURORA's edge-tier diagnosis and safety-gate behavior.

\section{Results and Discussion}
\label{sec:Results}

\subsection{Simulation setup}
\label{subsec:simulation}
To evaluate the proposed AURORA framework, we modeled a computing continuum environment and implemented it as a Python-based simulator. The environment emulates telemetry from a resource-constrained video-streaming edge device, such as a Raspberry Pi 4, and generates runtime state vectors comprising CPU utilization, memory usage, transmission delay, frame rate, and network throughput. Nominal operating states are sampled around predefined SLO thresholds using Gaussian perturbations, after which controlled grey-failure conditions are injected into selected telemetry dimensions. The evaluation includes 30,006 independent Monte Carlo trials, with 10,002 trials per agent variant (3,334 per fault type). In each trial, an agent observes the synthesized telemetry state, performs a diagnosis, and either executes a recovery action or abstains according to its decision logic. All experimental parameter values are listed in Table~\ref{tab:parameters}. The SLO thresholds $\phi_1$--$\phi_5$ were grounded in the physical constraints of the target hardware and stream format. The ranking weights $w_1$--$w_2$ and structural priors $r(C)$ were assigned by causal evidence strength, with the do-calculus contrast weighted most heavily, observed severity as secondary evidence, and a structural bias toward source nodes over downstream mediators. The Gate~1 base threshold requires the top hypothesis to carry substantially more posterior mass than the uniform-prior baseline, the floor is set at the point where the top hypothesis becomes less probable than all alternatives combined, and the decay rate is sized so the threshold reaches the floor when only a few SLO predicates are simultaneously violated. The Gate~2 threshold was selected at the empirical boundary between VFE distributions of safe and unsafe fault classes. The implementation and simulation scripts are publicly available for reproducibility.\footnote{\href{https://github.com/SuviDeSilva94/AURORA}{https://github.com/SuviDeSilva94/AURORA}}

The proposed AURORA framework was subjected to a dynamic, synthetically generated stream of ``grey faults.'' These faults are characterized by their ambiguous symptoms, which often trigger destructive cascading responses in traditional automated systems. The three fault classes injected during the trials were: 
\begin{itemize}
    \item \emph{Network Bandwidth:} Characterized by elevated delays and collapsed throughput.
    \item \emph{Anomalous Processing:} Characterized by processing starvation and rapid SLO violations.
    \item \emph{Memory Exhaustion:} Characterized by compounding RAM exhaustion without immediate bandwidth impacts.
\end{itemize}

\begin{table}[t]
\centering
\caption{Experimental parameters.}
\label{tab:parameters}
\begin{tabular}{llp{4.2cm}}
\toprule
\textbf{Parameter} & \textbf{Value} & \textbf{Description} \\
\midrule
\multicolumn{3}{l}{\textit{SLO Thresholds}} \\
$\phi_1$ & 33\,ms   & Network delay threshold \\
$\phi_2$ & 85\,\%   & CPU saturation threshold \\
$\phi_3$ & 85\,\%   & Memory saturation threshold \\
$\phi_4$ & 35\,fps  & Camera over-rate threshold \\
$\phi_5$ & 1.6\,MB/s & Bandwidth shortage threshold \\
\midrule
\multicolumn{3}{l}{\textit{Root-Cause Ranking Weights (Eq.~\ref{eq:ranking})}} \\
$w_1$ & 0.65 & Causal contrast weight \\
$w_2$ & 0.25 & Observation severity weight \\
$w_3$ & 0.10 & Direct-parent weight \\
$r(\cdot)_\text{root}$     & $+0.15$ & Root-node bonus \\
$r(\cdot)_\text{mediator}$ & $-0.10$ & Mediator-node penalty \\
\midrule
\multicolumn{3}{l}{\textit{Gate~1: Posterior Certainty (Eq.~\ref{eq:scaling})}} \\
$\tau_\text{base}$ & 0.70 & Base certainty threshold \\
$\tau_\text{min}$  & 0.50 & Minimum certainty floor \\
$\lambda$          & 0.15 & Threshold decay rate \\
\midrule
\multicolumn{3}{l}{\textit{Gate~2: VFE Safety}} \\
$\mathcal{F}_\text{th}$ & 3.85 & VFE execution threshold \\
\bottomrule
\end{tabular}
\end{table}

\subsection{Baselines}
\label{subsec:baselines}
AURORA's performance was formally evaluated against two control baselines representing heuristic and ungated inference-based approaches to edge orchestration:
\begin{itemize}
    \item \textbf{Rule-Based Agent:} A static, heuristic-driven model mapping observed symptoms to predefined actions via rigid thresholds (e.g., if CPU $>$ 80\%, execute restart). It possesses no causal reasoning capabilities.
    \item \textbf{AIF (No Gate) Agent:} A standard AIF model that utilizes BNs for root cause isolation but lacks the epistemic safety bounds. It natively executes the mitigation associated with the highest posterior probability without evaluating the VFE limits.
\end{itemize}

\subsection{Numerical Results}
\label{subsec:results}
\begin{figure*}[t]
    \captionsetup[subfigure]{justification=centering}
    \centering    
    \begin{subfigure}{0.25\textwidth}\centering
      \includegraphics[width=\linewidth]{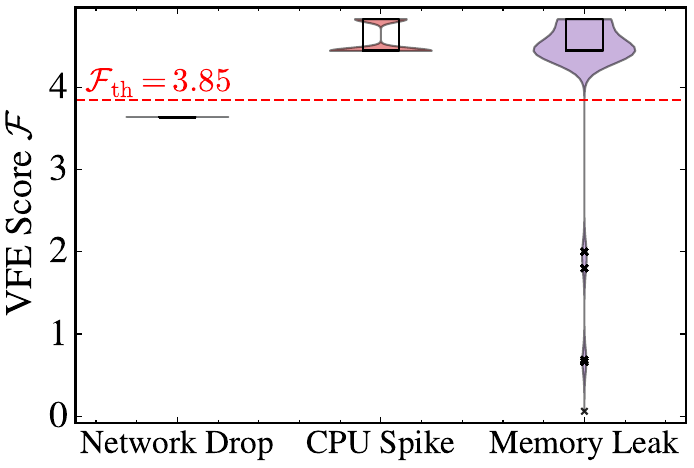}
      \caption{VFE per fault class \\(AURORA).}
      \label{fig:dist_vfe}
    \end{subfigure}\hfill
    \begin{subfigure}{0.25\textwidth}\centering
      \includegraphics[width=\linewidth]{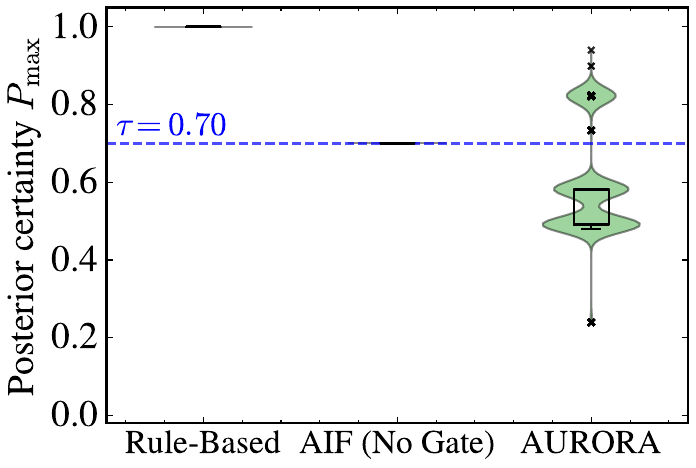}
      \caption{Posterior certainty $P_{\max}$ \\per agent.}
      \label{fig:dist_cert}
    \end{subfigure}\hfill
    \begin{subfigure}{0.25\textwidth}\centering
      \includegraphics[width=\linewidth]{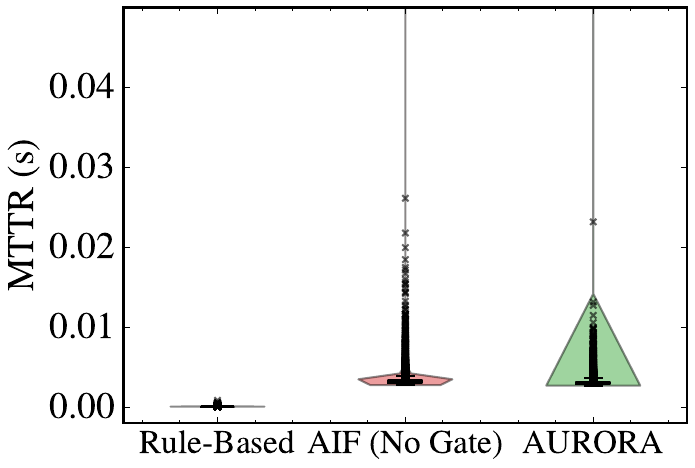}
      \caption{Decision latency \\per agent.}
      \label{fig:dist_mttr}
    \end{subfigure}\hfill
    \begin{subfigure}{0.25\textwidth}\centering
      \includegraphics[width=\linewidth]{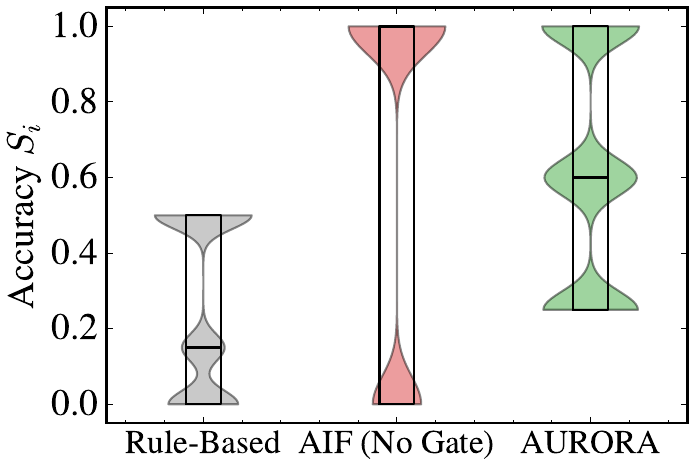}
      \caption{Per-trial outcome score $S_i$ \\per agent.}
    \end{subfigure}
    \caption{Per-trial distributions over the 10,002-trial Monte Carlo sweep. Each violin estimates the underlying distribution; the inner box reports the median and inter-quartile range with Tukey $1.5\!\times\!\text{IQR}$ whiskers, with outliers as $\times$ markers. Flat distributions reflect 
    constant-by-design agent behavior: network-drop trials produce a single VFE value ($\mathcal{F} \approx 3.64$); Rule-Based certainty is fixed at $1.00$ and AIF~(No~Gate) clips at $\tau = 0.70$. }
    \label{fig:distributions}
\end{figure*}

Table~\ref{tab:results} reports the aggregate metrics for the 
three agent architectures over the 30,006-trial Monte Carlo evaluation. Fig.~\ref{fig:distributions} further shows the per-trial distributions of the continuous metrics, including VFE score, decision latency, posterior certainty, and outcome score. Together, the table and figure capture both summary performance and variability across the i.i.d. fault samples.                                   

\begin{table*}[t]
\caption{Aggregated 
  experimental results (10,002 trials per agent)..}
\begin{center}
\resizebox{\textwidth}{!}{%
\begin{tabular}{@{}lcccccccc@{}}                                              
  \toprule                                                  
  \textbf{Agent} & \textbf{Repair Acc.} & \textbf{95\% CI} & \textbf{Resolution}
   & \textbf{Destructive} & \textbf{95\% Wilson CI} & \textbf{Abstention} &     
  \textbf{95\% Wilson CI} & \textbf{MTTR (s)} \\                                
  \midrule                                                                      
  Rule-Based & 26.2\% & [25.3\%, 27.1\%] & 46.2\% & 33.3\% & [32.4\%, 34.3\%] &
  0.0\% & [0.0\%, 0.04\%] & 0.0001 \\                                           
  AIF (No Gate) & 66.7\% & [65.7\%, 67.6\%] & 66.7\% & 33.3\% & [32.4\%, 34.3\%]
   & 0.0\% & [0.0\%, 0.04\%] & 0.003 \\                                         
  \textbf{AURORA} & \textbf{62.0\%} & \textbf{[61.4\%, 62.6\%]} & 34.1\% &
  \textbf{0.0\%} & \textbf{[0.00\%, 0.04\%]} & \textbf{65.9\%} &                
  \textbf{[65.0\%, 66.8\%]} & \textbf{0.003} \\             
  \bottomrule                                                                   
  \end{tabular} 
}
\label{tab:results}
\end{center}
\end{table*}

\subsubsection{
Destructive Action Rate}
An important metric of success for edge autonomy is preventing actions that exacerbate system instability. In the event of an ambiguous fault (such as misclassifying a network drop as a CPU failure), executing a container restart constitutes a destructive action.
AURORA successfully eliminated all destructive actions across   
  the entire 10,002-trial subset, validating the efficacy of the dual-gated     
  execution mechanism.

Crucially, because the framework successfully intercepted all deceptive anomalies, 3{,}411 out of 3{,}411 executed mitigations were diagnostically 
  correct (Fig. \ref{fig:abstention_triggers}), yielding a 100\% action precision rate for actions that passed the   
  dual gate.

\subsubsection{
Resolution Rate}
The resolution rate measures the percentage of trials where the agent successfully intervened and resolved the fault. Formally, it is defined as $(N_{\text{resolved}}\mathbin{/}N_{\text{total}}) \times 100$, where $N_{\text{resolved}}$ represents the count of true-positive mitigations that successfully recovered SLO compliance.

While AURORA possesses the lowest raw resolution rate, this is an 
expected
outcome of its safety architecture. The system is mathematically constrained from guessing under high uncertainty.
\subsubsection{
Abstention Rate}
The abstention rate defines the frequency at which the agent explicitly refuses to act due to high epistemic surprise or insufficient causal confidence.

The baseline agents act blindly, resulting in their high destructive action rates.  AURORA's 65.9\% abstention rate acts as a safety buffer, pausing edge       
   execution and offloading the ambiguous payload to a fog-tier orchestration 
   node for human-in-the-loop or heavy-compute evaluation.                     
                                                           
\begin{figure}[t]                                                                    
        \centering     
        \includegraphics[width=\columnwidth]{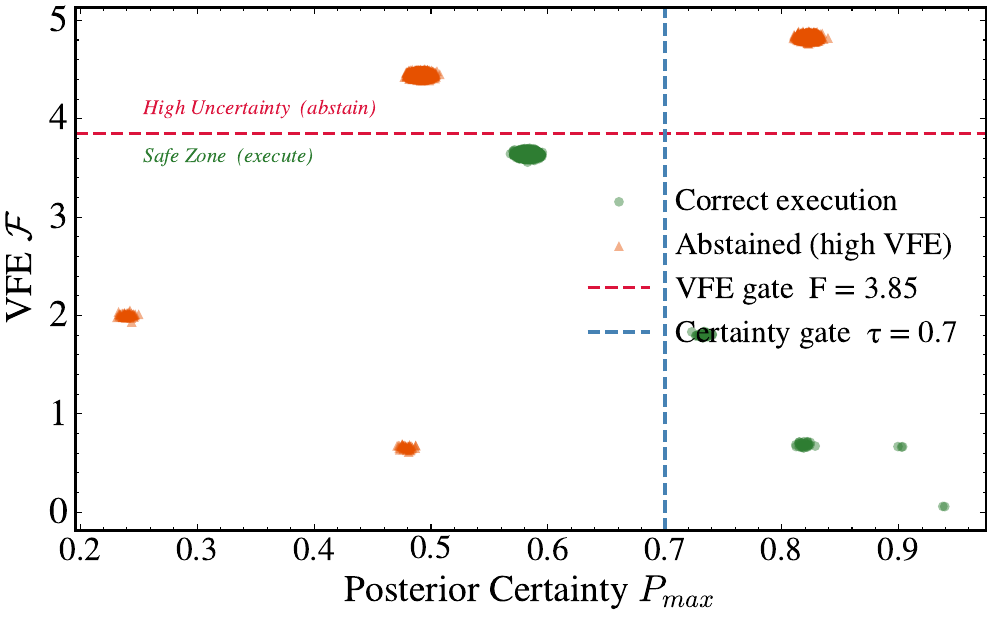}
        \caption{
        Safety gate decision space for AURORA across all 10,002 trials. Each point is one trial plotted by posterior certainty $P_{\max}$ (x-axis) and VFE score $\mathcal{F}$ (y-axis). Marker shape encodes fault type. The bottom right region marks the joint admissible zone ($P_{\max}\ge 0.70$ and $\mathcal{F} < 3.85$).
        }                                                                
        \label{fig:safety_gate_scatter}
      \end{figure}

\begin{figure*}[t]
    \captionsetup[subfigure]{justification=centering}
    \centering    
    \begin{subfigure}{0.25\textwidth}\centering
      \includegraphics[width=\linewidth]{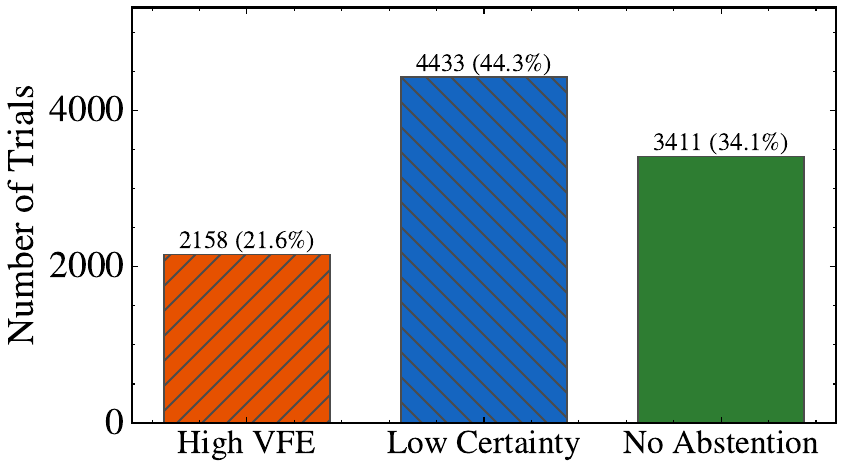}
      \caption{}
      \label{fig:abstention_triggers}
    \end{subfigure}\hfill
    \begin{subfigure}{0.25\textwidth}\centering
      \includegraphics[width=\linewidth]{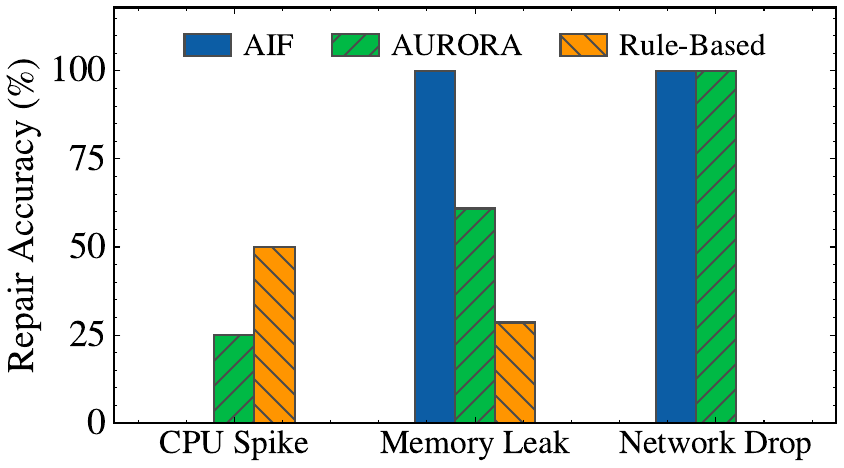}
      \caption{}
      \label{fig:fault_accuracy}
    \end{subfigure}\hfill
    \begin{subfigure}{0.25\textwidth}\centering
      \includegraphics[width=\linewidth]{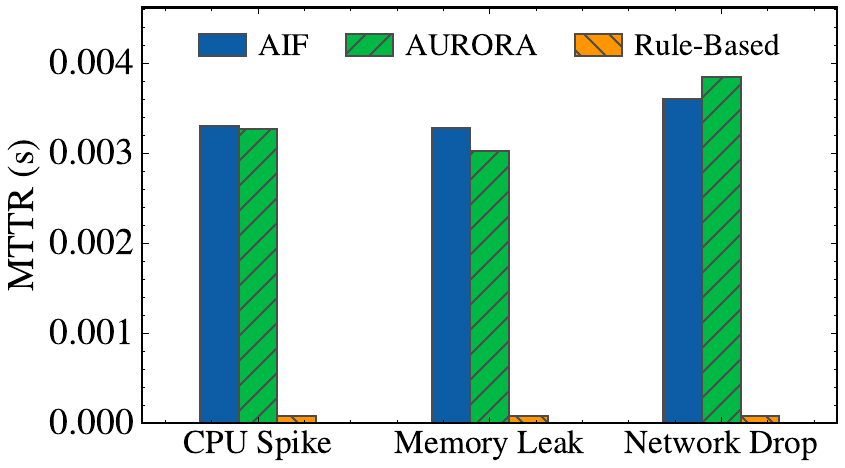}
      \caption{}
      \label{fig:fault_mttr}
    \end{subfigure}\hfill
    \begin{subfigure}{0.25\textwidth}\centering
      \includegraphics[width=\linewidth]{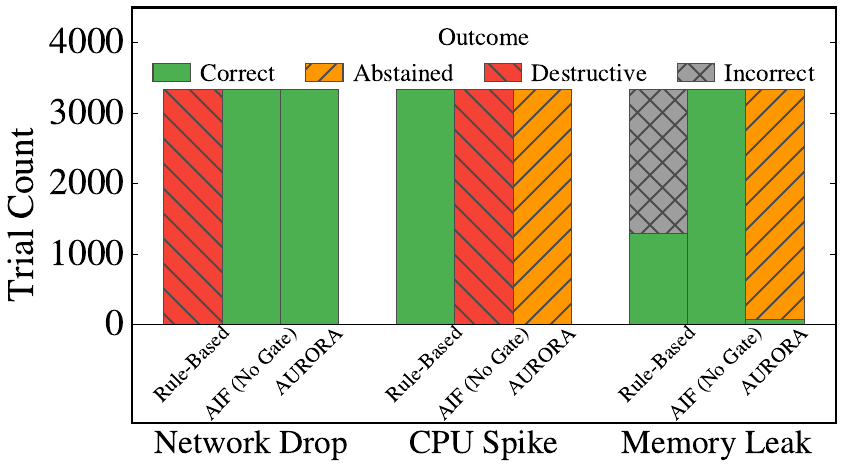}
      \caption{}
      \label{fig:outcome_composition}
    \end{subfigure}
    \caption{(a) AURORA's 10,002 trial outcomes by gate decision. Both safety
   gates contribute: the Posterior Certainty Gate fires on 4{,}433 trials       
  (44.3\%), the VFE Safety Gate on 2{,}158 trials (21.6\%), with 3{,}411 trials 
  passing both gates and executing locally.  (b) and (c) per-fault repair accuracy and decision latency over 10,002 trials per agent, respectively. (d) Per-fault outcomes across 3,334 trials per fault class per agent.}
\end{figure*}

   Fig.~\ref{fig:safety_gate_scatter} plots each trial in the joint $(P_{\max},
    \mathcal{F})$ decision space and reveals the precise mechanism behind this 
   rate. Of 10{,}002 trials, the Posterior Certainty Gate fires on 4{,}433     
   trials ($44.3\%$) and the VFE Safety Gate fires on 2{,}158 trials 
   ($21.6\%$). The two gates overlap on a subset and together account for
   $100\%$ of AURORA's abstentions (Fig.~\ref{fig:abstention_triggers}). Both
   gates are operationally active under the studied fault distribution. The dual-gated mechanism
   is therefore not redundant; each gate contributes distinct trials that the
   other does not catch.

Fig.~\ref{fig:outcome_composition} decomposes each agent's performance into per-fault outcome categories, revealing the precise mechanism behind the aggregate numbers in Table~II. The contrast between AIF~(No~Gate) and AURORA on CPU Spike faults is particularly illustrative: both agents receive identical telemetry, yet AURORA's VFE gate converts potential destructive actions into safe fog-tier offloads.    

\subsubsection{Accuracy}
  When evaluating the mean trial score accuracy, the framework assigns partial credit for safe isolation and catastrophic penalties for destructive interventions. Formally, the trial score $S_i$ is defined piecewise as $+1.0$ for a correct mitigation, $+0.6$ for a safe abstention on a recoverable fault class, $+0.25$ for a safe abstention on an unrecoverable fault class, $+0.15$ for an incorrect but benign action, and $0.0$ for a destructive action. The overall accuracy is defined as $\frac{1}{N}\sum_{i=1}^{N} S_i$. Under this score map, the 3{,}411 correct AURORA mitigations contribute               
  $3{,}411\cdot1.0$, the recoverable-class abstentions on memory-leak trials 
  contribute ${\sim}3{,}257\cdot0.6$, and the unrecoverable-class abstentions on
   CPU-spike trials contribute ${\sim}3{,}334\cdot0.25$, summing to $0.620$, as demonstrated in Table~\ref{tab:results}. Fig. \ref{fig:fault_accuracy} illustrates per-fault accuracy across the agents.

  AURORA maintains diagnostic accuracy on par with unbounded AIF, demonstrating that the safety constraints do not fundamentally impair the framework's ability to understand the system state.

\subsubsection{Mean Time to Repair}
Computational latency is a severe bottleneck on edge nodes.

As detailed in Fig. \ref{fig:fault_mttr}, despite the mathematical overhead of calculating Bayesian posteriors, Markov blankets, and VFE constraints, AURORA maintained an average Mean Time to Repair (MTTR) of $\sim$0.003 seconds. This confirms the computational feasibility of deploying advanced causal mechanisms natively on resource-constrained hardware.

\subsection{Discussion and Summary}
\label{subsec:discussion}
The 
Fig.~\ref{fig:dist_vfe} provides the mechanistic explanation for AURORA's abstention behavior. Network-drop trials generate $\mathcal{F}$ scores tightly below the gate and execute correctly; CPU-spike trials generate uniformly high $\mathcal{F}$ and always abstain; memory-leak trials exhibit a bimodal distribution driven by stochastic fault severity. Together with Fig. \ref{fig:dist_cert}, this confirms that the two gates measure orthogonal forms of uncertainty: a BN can be simultaneously \emph{confident} (high $P_{\max}$) and \emph{epistemically surprised} (high $\mathcal{F}$) when its causal model mismatches the observed telemetry, and only the VFE gate captures this distinction.

The results confirm a fundamental trade-off: the more autonomously an agent resolves faults, the more destructive actions it risks, and the safer it acts, the fewer faults it resolves.
While the standard ungated AIF model successfully resolved 66.7\% of anomalies, it did so at an unacceptable cost of a 33.3\% destructive action rate, which in a real-world computing continuum could lead to catastrophic cascading network failure. 
In contrast, AURORA trades a ${\sim}32.6$ percentage-point reduction in autonomous blind resolution (as shown in Fig.~\ref{fig:abstention_triggers}) for a $100\%$ elimination of destructive actions. Through the ``intentional abstention'' paradigm, the framework shows that recognizing and bounding uncertainty is more valuable than forced autonomy in unpredictable grey-fault conditions.

Although AURORA demonstrates the feasibility of uncertainty-aware micro-agent resilience at the edge, several limitations remain. The current SLO predicates, recovery actions, action-to-patch mappings, and gate thresholds are manually configured and should be calibrated using representative ground-truth traces for deployment-specific operation. Additionally, the Monte Carlo environment does not fully capture physical edge--fog effects such as telemetry noise, actuation delay, clock drift, contention, and signaling overhead as the number of micro-agents and monitored variables increases.

\section{Conclusions}\label{sec:Conclusion}

This paper presented AURORA, an uncertainty-aware micro-agent framework for grey failure diagnosis and mitigation on resource-constrained edge devices in the computing continuum. AURORA performs localized causal reasoning without requiring full continuum-wide state evaluation, and its dual-gated execution mechanism permits remediation only when posterior causal confidence is sufficient, and VFE remains bounded; otherwise, unresolved cases are escalated to the fog tier. Monte Carlo evaluation across network collapse, memory leak, and processing spike scenarios showed that AURORA eliminated destructive interventions, achieving a $0.0\%$ destructive action rate compared with $33.3\%$ for rule-based and ungated AIF baselines, while maintaining $62.0\%$ repair accuracy and approximately 3ms MTTR. These results confirm that uncertainty-aware abstention can improve the safety of autonomous edge remediation without imposing significant latency overhead. 
Future work will focus on deploying AURORA on a physical edge--fog testbed to evaluate performance under realistic conditions and assess scalability under live grey-failure conditions.

\bibliographystyle{ieeetr}
\bibliography{ref}
\balance
\end{document}